\documentclass[a4paper]{article}
\usepackage{graphicx}
\usepackage{amsmath,amssymb}

\date{}
\title{Contour Dynamics for One-Dimensional Vlasov-Poisson Plasma with the Periodic Boundary}
\author{Hiroki Sato$^*$, T.-H. Watanabe and Shinya Maeyama\\ %ここに著者名を書きます。
Department of Physics, Nagoya University, Furo-cho, Nagoya, 4646-8602, Japan
}
\begin{document}
\maketitle
\begin{abstract}
We revisit the contour dynamics (CD) simulation method which is applicable to large deformation of distribution function in the Vlasov-Poisson plasma with the periodic boundary, where contours  of distribution function are traced without using spatial grids. Novelty of this study lies in application of CD to the one-dimensional Vlasov-Poisson plasma with the periodic boundary condition. A major difficulty in application of the periodic boundary is how to deal with contours when they cross the boundaries. It has been overcome by virtue of a periodic Green's function, which effectively introduces the periodic boundary condition without cutting nor reallocating the contours. The simulation results are confirmed by comparing with an analytical solution for the piece-wise constant distribution function in the linear regime and a linear analysis of the Landau damping. Also, particle trapping by Langmuir wave is successfully reproduced in the nonlinear regime.
\end{abstract}

\section{Introduction}
 Kinetic equations for plasma dynamics describe many interesting physical phenomena, but are generally difficult to be solved analytically or numerically. For example, a long term nonlinear evolution of the distribution function is not yet fully understood even in the one-dimensional Vlasov-Poisson system. Three types of simulation methods for kinetic plasma are widely known, such as Lagrangian, semi-Lagrangian, and Eulerian methods. The Particle-In-Cell (Lagrangian) method has a problem of numerical noise, while resolution of the Vlasov method (Eulerian) is limited by the grid size. Indeed, it is shown that, in the Vlasov simulation of the nonlinear Landau damping,  fine structures of the distribution function continue to grow in phase space and  are stretched exponentially in time, increasing numerical errors (Ref.~\cite{watanabe}).\par
The water-bag model, which assumes a piece-wise constant distribution function $(f)$, has been studied for the Vlasov-Poisson plasma (Refs~.{\cite{Berk1967,depackh1962water,morel2008}) since 1960s, and successfully resolved stretching and strong deformation of $f$ in the phase space $(x,v)$. In 1979, as a generalization of the water-bag model, contour dynamics (CD) method is introduced by  Zabusky, Hughes, and Roberts (Ref.~\cite{zabusky1979}) for solving inviscid and incompressible fluid motions in the two-dimensional configuration space $(x,y)$. The CD method employs nodes on each contour of which motion is given by calculating line integrals of the Green's function along contours. Because CD employs no spatial grid but nodes on contours (Lagrangian), the numerical resolution is not limited by spatial grids, which makes the CD method tough against large deformation of vorticity (Ref~.\cite{review}). \par
In this paper, we revisit the CD method and apply it to the Vlasov-Poisson plasma with the periodic boundary condition. Although the basic idea of the CD method stems from the water-bag model, there has been a few applications to the Vlasov-Poisson plasma. The CD method differs from the water-bag model as no spatial grid is used in the former in solving Poisson equation (Ref.~\cite{Berk1967}). Although a modern implementation of the water-bag method by Colombi \& Touma (Ref.~\cite{colombi2008vlasov}) did not use spatial grids, the application is limited to a system with no spatial boundary. Novelty of the present paper lies in application of the CD method to the Vlasov-Poisson plasma with the periodic boundary, where we consider time development of contours of the distribution function without using spatial grids. The simulation results are confirmed by comparing with an analytical solution for the piece-wise constant distribution function in the linear regime and a linear analysis of the Landau damping. Furthermore, the particle trapping by Langmuir waves is successfully reproduced in the nonlinear regime. Here, it should be remarked that no contour surgery (Ref.~\cite{Dritschel1988}) nor node redistributions (Ref.~\cite{Vosbeek}) is employed in numerical simulations in this paper, because we focus on validity of our implementation for the periodic boundary condition.\par
This article is organized as follows. After a brief introduction to the CD in Section~\ref{ContourDynamics}, application to the Vlasov-Poisson system with the periodic boundary is described in Section~\ref{AVPS}. Validity of the CD method is confirmed by comparing the simulation results with the analytical solution for the piece-wise constant distribution function in Section~\ref{benchmark}. A bench mark test for the linear Landau damping is described in Section~\ref{sL1}.  Application to the nonlinear Landau damping is shown in Section~\ref{sL2}. Finally, we summarize the results in Section~\ref{SumAndCon}.
\clearpage

\section{Contour Dynamics}\label{ContourDynamics}
Zabusky, Hughes, and Roberts have proposed contour dynamics algorithm for the Euler equation of fluid dynamics in two dimensions (Ref.~\cite{zabusky1979}). The governing equations are
\begin{align}
& \frac{D\omega}{Dt} = \frac{\partial\omega}{\partial t} + u\frac{\partial \omega}{\partial x} + v\frac{\partial\omega}{\partial y} = 0,\label{omegav}\\
&\nabla^2 \psi = \frac{\partial^2\psi}{\partial x^2} + \frac{\partial^2 \psi}{\partial y^2} = -\omega,\label{psip}\\
& u = \frac{\partial \psi}{\partial y}, v= -\frac{\partial \psi}{\partial x},\notag\\
& \omega = -\frac{\partial u}{\partial y} + \frac{\partial v}{\partial x},
\end{align}
where $\psi$ is the stream function and $\omega$ means the vorticity. Time development of the vorticity is calculated by tracing motions of contours of the piece-wise constant vorticity distribution.The flow velocity is given by the line integrals of the Green's function on the contours, such that
\begin{align}
&(u,v) = \left(\frac{\partial \psi}{\partial y},-\frac{\partial \psi}{\partial x}\right) = \displaystyle{\sum_{m}}(\Delta \omega_m)\oint_{C_m} G(\xi,\eta;x,y)dr'_m,
\end{align}
where $m$ is a label of contours, $C_m$ is the contour labeled by $m$, and $\Delta \omega_m$ is a jump of vorticity when crossing the contour $C_m$ inward. The two-dimensional Green's function, $G(x,y;\xi,\eta)$, is given as
\begin{align}
&G(x,y;\xi,\eta) = -\frac{1}{2\pi}\log \sqrt{(x-\xi)^2 + (y-\eta)^2}.
\end{align}
The contours are discretized by nodes with label $n$, and motion of the contours is determined by solving the Hamilton equations of the node points (Fig.~\ref{cd}).  Each contour represents a constant vorticity line, and is advected by an incompressible flow as given in Eq.~(\ref{omegav}). The incompressibility also guarantees conservation of volumes surrounded by each contour.
\begin{figure}[h]
\centering
\includegraphics[scale=.5]{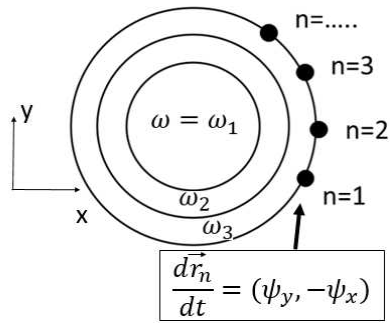}
\caption{In contour dynamics, Motion of the contours are determined by solving the Hamilton equations of the node points}
\label{cd}
\end{figure}
\clearpage

\section{Appication to Vlasov-Poisson system}\label{AVPS}
\subsection{Basic scheme}
Here, we consider application of the CD method to the Vlasov-Poisson system with the periodic boundary. The normalized Vlasov-Poisson equations are
\begin{align}
&\frac{\partial f}{\partial \tau}+ v\frac{\partial f}{\partial x} +a\frac{\partial f}{\partial v}= 0,\label{bvpvlasov}\\
&a=\frac{\partial \phi}{\partial x},\\
&-\nabla^2 \phi = 1 - \int^{\infty}_{-\infty} f(x,v) dv =: F(x)\label{bvppoisson},
\end{align}
where  $f$ is the distribution function of electrons, while stationary background ions are assumed. The particle density is normalized so that $\int^{L/2}_{-L/2}dx\int^{\infty}_{-\infty}dv f(x.v) = L$, where $L$ denotes  the system length. The periodic boundary conditions at $x=\pm L/2$ are given by
\begin{align}
&\lim_{\epsilon \rightarrow -0}\phi\left(\frac{L}{2}+\epsilon\right) = \lim_{\epsilon' \rightarrow +0}\phi\left(-\frac{L}{2}+\epsilon'\right),\label{phibc1}\\
&\lim_{\epsilon \rightarrow -0}\phi'\left(\frac{L}{2}+\epsilon\right) = \lim_{\epsilon' \rightarrow +0}\phi'\left(-\frac{L}{2}+\epsilon'\right)\;(\text{quasi neutrality}),\label{phibc2}
\end{align}
The Liouville's theorem and  Eq. (\ref{bvpvlasov}) guarantee the volume conservation and  $\frac{df}{d\tau} = 0$, which are required for contour dynamics method. In order to implement the CD, we employ the Green's function, $G$, that satisfies
\begin{align}
&\nabla^2 G\left(x;\xi\right) = \frac{1}{L} - \delta\left(x-\xi\right),\label{gpoisson}\\
&\lim_{\epsilon \rightarrow -0}G\left(\frac{L}{2}+\epsilon\right) = \lim_{\epsilon' \rightarrow +0}G\left(-\frac{L}{2}+\epsilon'\right), \label{gbc1}\\
&\lim_{\epsilon \rightarrow -0}G'\left(\frac{L}{2}+\epsilon\right) = \lim_{\epsilon' \rightarrow +0}G'\left(-\frac{L}{2}+\epsilon'\right).\label{gbc2}
\intertext{Solving  Eqs. (\ref{gpoisson}), (\ref{gbc1}), and (\ref{gbc2}) to obtain $G$ (Ref.~\cite{Green}), one finds}
 &G\left(x;\xi\right)= \frac{1}{2L}\left(|x-\xi|-\frac{L}{2}\right)^2.  \label{Gfunct}\\
\intertext{Therefore,}
\phi(x) &= \int^{\frac{L}{2}}_{-\frac{L}{2}} G\left(\xi;x\right)F\left(\xi\right)d\xi +\text{const}\text{\;\;\;\;for $x\in\left(-\frac{L}{2},\frac{L}{2}\right)$}\label{phi}.\\
\intertext{The acceleration of each particle at $x$ is given by the CD representation,}
&a\left(x\right) = \displaystyle{\sum_m^{N_m}} \Delta f_m \oint_{c_m} G\left(\xi;x\right)dv,\label{a=og}
\end{align}
where $N_m$ is the number of contours, $C_m$ is a contour labeled by $m$ and $\Delta f_m$ is the jump of distribution function when crossing the contour $C_m$ inward. We discretize the contours with nodes labeled counterclockwise by $n$ and connect nodes with straight line segments. Then, Eq. (\ref{a=og}) breaks down into
\begin{align}
&a\left(x\right) =   \displaystyle{\sum_m} \Delta f_m \displaystyle{\sum_n}\frac{v_{n+1}- v_n}{2}\left( \frac{w_n^2(x)}{L} + \frac{\delta_n^2}{12L} - I_n(x) \right),\label{pa=og}
\intertext{where}
&I_n(x) = \begin{cases}|w_n(x)| &\text{for\;\;}  |w_n(x)| \geq \frac{\delta_n}{2}\\
\frac{w_n^2(x)}{\delta_n} + \frac{\delta_n}{4} &\text{for\;\;}|w_n(x)| < \frac{\delta_n}{2}
\end{cases}\text{\;\;with\;\;}w_n(x) := x-\frac{x_{n+1}+x_n}{2}\text{\;\;and\;\;} \delta_n := |x_{n+1} - x_{n}|.
\end{align}
Although $n$ is a function of the label of contours ($m$), we use the notation of $n$ = $n(m)$ for simplicity. Equation of motion of each node point labeled by $i$ is given by
\begin{align}
\frac{d^2x_i}{d\tau^2} = a\left(x=x_i\right) =   \displaystyle{\sum_m} \Delta f_m \displaystyle{\sum_n}\frac{v_{n+1}- v_n}{2}\left( \frac{w_n^2(x)}{L} + \frac{\delta_n^2}{12L} - I_n(x) \right).
\end{align}
For the time integration, we use the leap-frog scheme with the time step size $\Delta \tau$ = 0.01 in the all simulations shown below.
\subsection{Implementation of the periodic boundary}
A difficulty of implementation arises in the CD method with the periodic boundary, when a node$(x_n,v_n)$ moves across the boundaries and comes into the simulation box from the another side. Straightforwardly, we may cut the contour at the boundary and reallocate a node point $(x_n,v_n)$ as
\begin{align}
&x_n \notin \left(-L/2,L/2\right) \Rightarrow \begin{cases}
x_n \mapsto x_n - L/2 &;\text{if }L/2<x_n\\
x_n \mapsto x_n + L/2 &;\text{if }x_n<-L/2\\
\end{cases}
\end{align}
and make interpolation of points $\left(\tilde{x_s},\tilde{v_s}\right)$ and $\left(\tilde{x_t},\tilde{v_t}\right)$ on the boundary between $\left(x_{n+1},v_{n+1}\right)\notin\left(\frac{-L}{2},\frac{L}{2}\right)$ and $\left(x_{n},v_{n}\right)\in\left(\frac{-L}{2},\frac{L}{2}\right), $
\begin{align}
&\left(\tilde{x_s},\tilde{v_s}\right) = \begin{cases}
\left(\frac{L}{2},\frac{v_{n+1} - v_{n}}{x_{n+1} - x_{n}}\left(\frac{L}{2}-x_{n+1}\right) + v_{n+1}\right)&;\text{if }L/2\leq x_{n+1}\;\;\;\cdots A\\
\left(\frac{-L}{2},\frac{v_{n+1} - v_{n}}{x_{n+1} - x_{n}}\left(\frac{-L}{2}-x_{n+1}\right) + v_{n+1}\right)&;\text{if }x_{n+1}\leq -L/2\;\;\;\cdots B\\
\end{cases}\label{xs}\\
&\left(\tilde{x_t},\tilde{v_t}\right) = \begin{cases}
\left(\frac{-L}{2},\frac{v_{n+1} - v_{n}}{x_{n+1} - x_{n}}\left(\frac{L}{2}-x_{n+1}\right) + v_{n+1}\right)&;\text{if }L/2\leq x_{n+1}\;\;\;\cdots A'\\
\left(\frac{L}{2},\frac{v_{n+1} - v_{n}}{x_{n+1} - x_{n}}\left(\frac{-L}{2}-x_{n+1}\right) + v_{n+1}\right)&;\text{if }x_{n+1}\leq -L/2\;\;\;\cdots B'\\
\end{cases}\label{xt}
\end{align}
for calculation of the line integrals (see fig.~\ref{case1}). here, we call this method the reallocation scheme. since we must know the sequence of the nodes for the cd method, the reallocated nodes complicate the computational algorithm. actually, we need to count how many times each node moved across the boundaries and to use the counts every time in calculation of $\frac{v_{n+1}- v_n}{2}\left( \frac{w_n^2}{l} + \frac{\delta_n^2}{12l} - i_n \right)$. however, it increases numerical costs and makes the code implementation complicated.
\begin{figure}[h] %option[],need{}
\centering
\includegraphics[width=8cm]{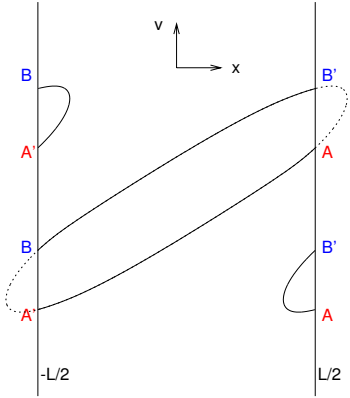}
\caption{Conventional implementation of the periodic boundary. Contours which get out of simulation box are cut and reallocated ($A, B, A'$ and $B'$ are defined on Eqs. (\ref{xs}) and (\ref{xt})).}
\label{case1}
\end{figure}
\par
In the following, we propose a novel scheme to implement the periodic boundary in CD, which is named a periodic Green's function method. Because we consider the periodic problem; $f(a)=f(a+L),$ $\phi(a)=\phi(a+L),$ and $\phi'(a) = \phi'(a+L)$. Thus, it may be possible to eliminate the boundaries at $x = \pm\frac{L}{2}$, while extending the simulation box to $(-\infty,\infty)$ and imposing the periodicity to the Green's function, that is,
\begin{align}
&\forall x\in(-\infty,\infty),\;\;\;a(x) = \displaystyle{\sum_{m=1}^{N_m}}\Delta f_m\oint_{c_m} G'\left(\xi;x\right) dv'\label{moda=og},
\intertext{with}
&G'\left(\xi;x\right) = G\left(Mod\left(\xi-\left(x-L/2\right),L\right)+\left(x-L/2\right) ; x\right),\\
\intertext{where}
&Mod\left(a,b\right) = Min\left\{c\left(\geq0\right)|\exists r\in\mathbb{Z}, a = br+c\right\}.
\end{align}
In this way, we can avoid cutting or reallocation of the contours, but equivalently the contours feel the periodicity of the system through $G'$ instead of $G$ (See Fig.~\ref{case2}).  If $\forall n,|x_n - x_{n-1}| < \frac{L}{2}$ is satisfied, then (\ref{pa=og}) becomes
\begin{align}
 &a(x) =   \displaystyle{\sum_m} \Delta f_m \displaystyle{\sum_n}\frac{v_{n+1}- v_n}{2}\left( \frac{w_n'^2}{L} + \frac{\delta_n^2}{12L} - I'_n \right)\label{modpa=og},
\intertext{where}
&w_n'(x):= x+r\left(x_n,x\right)L - \frac{x_{n+1} + x_n}{2}, \\
&r\left(x_n,x\right) : = \frac{x_n - \left(x-\frac{L}{2}\right)-Mod\left(x_n-\left(x-\frac{L}{2}\right),L\right) }{L} = \lfloor\frac{x_n - \left(x - \frac{L}{2}\right)}{L}\rfloor, \notag\\
\intertext{and}
&\delta_n:= |x_{n+1} - x_n|,\;\;I'_n(x) = \begin{cases}|w_n'(x)| &\text{for\;\;}  |w_n'(x)| \geq \frac{\delta_n}{2}\\
\frac{w_n'^2(x)}{\delta_n} + \frac{\delta_n}{4} &\text{for\;\;}|w_n'(x)| < \frac{\delta}{2}\end{cases}.
\end{align}
Here, the periodicity is introduced not in $C_m$, but in $G'$. Therefore we do not need to count how many times each node moves across the boundaries, which makes a faster and simpler implementation. Our new implementation with the periodic Green's function $G'$ accelerate the computation speed
%about 1.5
 faster than that of the reallocation scheme. It owes to no cutting nor reallocation, which makes the code avoid many "if" branches.
\begin{figure}[h]
\centering
\includegraphics[width=8cm]{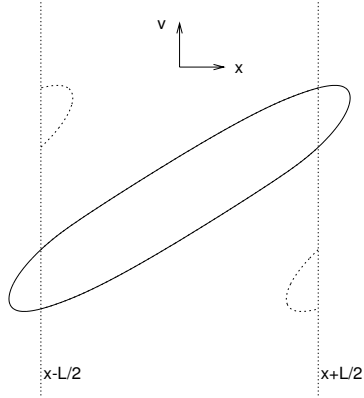}
\caption{Implementation of the periodic boundary using $G'= G(Mod(\xi-(x-L/2),L)+(x-L/2) ; x)$. The periodic boundary is effectively introduced without cutting nor reallocating the contours.}
\label{case2}
\end{figure}
\clearpage

\section{Benchmark Test for Piece-Wise Constant Distribution Function}\label{benchmark}
In order to check validity of our application, we consider the initial distribution function given by
\begin{align}
&f\left(x,v,\tau=0\right) = \displaystyle{\sum_{m=-N_m}^{N_m}}b_m U\left(v-v_m(x,0)\right),\label{linin}\\
&U(x) = \begin{cases}
1& 0<x\\
0&  x\leq 0
\end{cases},
\end{align}
where $b_m=-b_{-m}<0$ and $v_m(x,0) = v^0_m + v^1_m(x)$. Also $v^0_m = m\Delta v$ and $v^1_m(x)  = \alpha e^{ikx}$ with $\Delta v\in\mathbb{R}^{+}$ and $kL/2\pi\in\mathbb{N}$, where $\alpha \ll 1$ so that $|v^1_m|\ll |v^0_m|.$
This function Eq. (\ref{linin}) was also used to study the water-bag model (Ref.\cite{Ldawb}). As shown in Appendix, the linear dispersion relation is derived as
\begin{align}
&D(\omega) =1 +\displaystyle{\sum_{m>0}}\frac{2b_mv^0_m}{\omega^2 - \left(kv^0_m\right)^2}=0.\label{anaD}
\intertext{Solving the initial value problem analytically, the acceleration $a(x,\tau)$ is determined by means of $\omega_l$ which satisfies $ \;D(\omega_l) = 0 $ with $ kv^0_l < \omega_l < kv^0_{l+1}$,}
a(x,\tau) &=  Re\left[\displaystyle{\sum_{j>0}}\frac{\displaystyle{\sum_{m>0}}\left\{\left(-b_m\right)\alpha e^{i(kx - \frac{\pi}{2})}\Pi_{n(\neq m)>0}\left(\left(\omega_j\right)^2 - \left(kv^0_n\right)^2\right)\right\}}{k\Pi_{n(\neq j)>0}\left(\omega^2_j-\omega^2_n\right)}2\cos\omega_j \tau\right].\label{kaiseki}
\end{align}
%位相のプラマイは＋に修正したけどどうなのかな？？
We make comparison of a simulation result with Eqs.~(\ref{anaD}) and (\ref{kaiseki}). \par
For $N_m=2,\;b_1=b_2 = -1/6,\;\Delta v = 1.0,\;k = 1.0$ and $\alpha = 0.01$, Eqs.~(\ref{anaD}) and  (\ref{kaiseki}) lead $a(x=-1.4\pi,\tau) = A_1 \cos(\omega_1 \tau) + A_2 \cos(\omega_2 \tau)$ with $A_1= 0.00226,\;A_2 = 0.00409,\;\omega_1=1.13$ and $\omega_2 = 2.18.$ The numerical result of $a(x=-1.4\pi,\tau)$ shown in Fig.~\ref{anl2} agrees well with the analytical prediction, Eq.~(\ref{kaiseki}). It shows validity of our implementation of the CD method for the Vlasov-Poisson system with the periodic boundary~condition.

\begin{figure}[h]
\centering
\includegraphics[angle=-90,width=8cm]{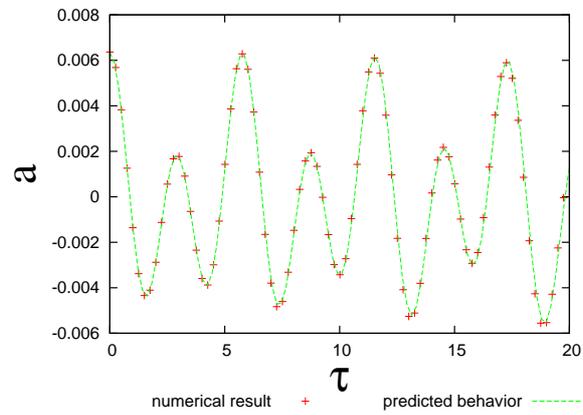}
\caption{Benchmark test1: Time evolution of $a = \partial \phi/\partial x$. Red points are simulation results and green line is predicted by the Eq.~(\ref{kaiseki}). $N_m=2,\;b_1=b_2 = -1/6,\;\Delta v = 1.0,\;k=1.0$ and $\alpha = 0.01$ are given  for initial distribution function (\ref{linin}).}
\label{anl2}
\end{figure}
\clearpage

\section{Benchmark test for the Landau Damping}
\subsection{Linear Landau Damping}\label{sL1}
We also verify our code for the linear Landau damping. We set the initial contour distribution as follows. The initial (continuous) distribution function $f$ is given by
\begin{align}
&f(x,v,\tau=0) = \frac{1}{\sqrt{2\pi}}\exp\left(-\frac{v^2}{2}\right)(1+ \alpha\cos(kx))\label{inid}.
\intertext{By means of a sequence $\{\Delta f_m\}^{N_{Max}}_{m=1}:\Delta f_m \in \mathbb{R}^+ $ and $\displaystyle{\sum^{N_{Max}}_{m=1}}{\Delta f}_m < \underset{x,v\in\mathbb{R}}{Max}\left\{f(x,v,\tau=0)\right\},$ we define $\tilde{f}$, a piece-wise constant approximation of $f$,}
&\tilde{f}(x,v,\tau=0) := \displaystyle{\sum^{N_m}_{m=1}} \Delta f_m I\left[\Sigma_{m'=1}^{m}\Delta f_{m'}< f(x,v,\tau=0)\right],\label{tilf}\\
\intertext{where}
&I\left[P(x,v,\tau)\right] = \begin{cases}1
& \text{if  $P(x,v,\tau)$ is true}\\
0& \text{otherwise}
\end{cases}
\end{align}
with a propositional function $P(x,v,\tau)$. However, Eq.~(\ref{tilf}) does not satisfy $\int^{\frac{L}{2}}_{-\frac{L}{2}}dx\int^{\infty}_{-\infty} \tilde{f}(x,v,t=0)dv = L$ because $\tilde{f}$ is a piece-wise constant function defined by means of contours of $f$. Therefore instead of $\tilde{f}$, we use $\tilde{f}'$ normalized by $(1+\epsilon)$,
\begin{align}
&\tilde{f}' = (1+\epsilon)\tilde{f}, \text{\;\;where\;\;}\epsilon =\frac{L}{\int^{\frac{L}{2}}_{-\frac{L}{2}}dx\int^{\infty}_{-\infty}\tilde{f}(x,v,t=0) dv}-1,
\end{align}
and thus $\Delta f'_m = (1+\epsilon)\Delta f_m$.\par %noting $Max\{\Delta f_m\}\rightarrow 0\Rightarrow\epsilon\rightarrow0$.\par
The simplest way of giving $\{\Delta f_m\}$ is $\Delta f_m=$constant. However, contour dynamics method does not require constant $\Delta f_m$, and non-uniform contour intervals have an advantage over the constant $\Delta f_m$ in approximation of $f$. In case with $\Delta f_m =$ constant, the contours are densely distributed where the velocity space gradient of $f$ is steep around $v\sim\pm v_{th}$ ($v_{th}$ means the thermal velocity), while no contour is found  for $|v| > 3v_{th}$ when we use 40 contours. It means that there is no particle in $|v|>3v_{th}$, while the super thermal particles can be included in the case of $\Delta f_m \neq$ constant. From the linear theory, real and imaginary parts of the eigenfrequency for $k=0.5$ are evaluated as $\omega_r = 1.4156$ and $\gamma = -0.1533$. Thus, the phase velocity is $\omega_r/k \sim 2.8v_{th}$. This is the reason why the high speed particle should be included in this application. Otherwise, many contours are necessary in the constant $\Delta f_m$ case in order to introduce contributions of the super thermal particles. Thus, we employ the non-uniform contour intervals of $\Delta f_m$ which is defined as
\begin{align}
\Delta f_m : = &\begin{cases}
f(x=x_0,v=V_m) & m = 1\\
f(x=x_0,v=V_m) - f(x=x_0,v=V_{m-1}) &2 \leq m \leq N_{Max}\\
 \end{cases},\label{delaf}
 \end{align}
with $d v \in\mathbb{R}^{+},N_{Max}\in\mathbb{N}$ and $V_m:=(N_{Max}+1-m)dv$. Since Eq.~(\ref{inid}) has the maximum at $x=0$, we set $x_0 =0$. In this application, we used 40 contours with $dv = 0.1$, where the contour spacing in $v$ is nearly constant covering the velocity space of $|v|<4v_{th}$. \par
A simulation result for the linear Landau damping  is presented in Fig.~\ref{lin}, where the initial distribution function is given by Eq.~({\ref{delaf}}) with $\alpha=0.01$ and $k=0.5.$ The simulation box size is $L=4\pi$. We also used 2000 nodes/contour. One clearly finds the linear damping rate of $\gamma = -0.153$ successfully reproduced by the present CD method.

\begin{figure}[h]
\centering
\includegraphics[angle=-90,width=8cm]{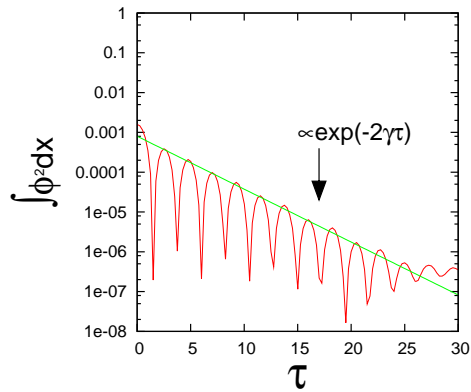}
.\caption{Time history of the quadratic integral of the electrostatic potential $\phi$ obtained from the simulation of the linear Landau damping. The damping rate $\gamma$ = -0.153 is successfully confirmed by the contour dynamics method.}
\label{lin}
\end{figure}
\clearpage

\subsection{Nonlinear Landau damping}\label{sL2}
Next, we consider a benchmark test for the nonlinear Landau damping (Ref.~\cite{watanabe}) with the initial distribution function in Eq.~(\ref{inid}) where $\alpha =0.5$ and $k=0.5$. We employ 40 contours and 8000 nodes/contour. It is noteworthy that intersections of contour lines are not observed till $\tau=30$ in the present simulation. It is, thus, appropriate not to use the node redistribution nor the contour surgery in the current test case. Fig.~\ref{nonlin} shows a snapshot of contour distribution in the phase space at $\tau = 30$, where the particle trapping by the Langmuir wave is successfully reproduced by the CD method. \par
\begin{figure}[h]
\centering
\includegraphics[angle=-90,scale=.5]{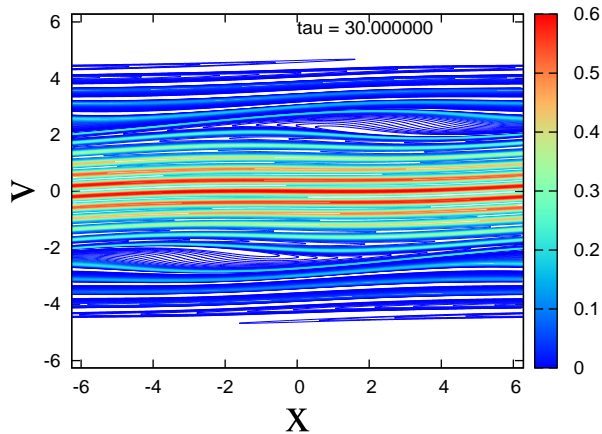}
\caption{Phase space structure of nonlinear Landau damping at $\tau = 30$. Particles trapped by waves are reproduced. The color bar represent the magnitude of $f$.}
\label{nonlin}
\end{figure}
Soundness of our implementation is also confirmed by conservation of energy, as shown in Figs~(\ref{nonlin_ene}) and (\ref{nonlin_eren}).
Total energy, $E_t = \frac{1}{2}\iint v^2 f dxdv + \frac{1}{2}\int |\frac{\partial \phi}{\partial x}|^2 dx$, is conserved with an error, $\epsilon_{E(\tau)}=|E_t(\tau) - E_t(0)|/E(0)$, less than $2.5\times10^{-5}$ for $\tau \leq 30$.\par
\begin{figure}[h]
\centering
\includegraphics[angle=-90,scale=.5]{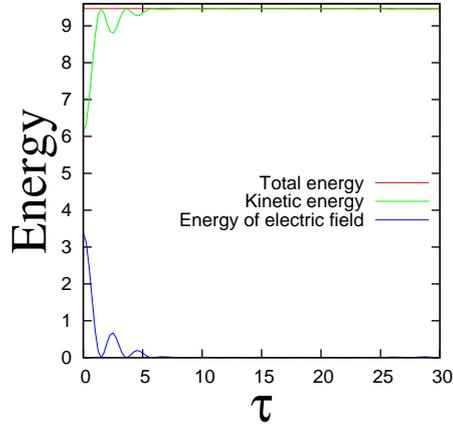}
\caption{Conservation of energy. Blue line represents energy of electric field, $E_\phi = \frac{1}{2}\int |\frac{\partial \phi}{\partial x}|^2 dx$ , Green kinetic energy, $E_k =\frac{1}{2}\iint v^2 f dxdv$, and Red Total energy, $E_t = E_\phi + E_k$.}
\label{nonlin_ene}
\end{figure}
\begin{figure}[h]
%\centering
\includegraphics[angle=-90,scale=.5]{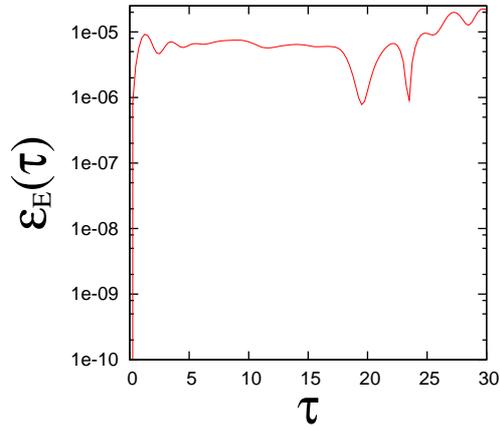}
\centering
\caption{The error found in total energy, $\epsilon_E(\tau)=|E_t(\tau) - E_t(0)|/E(0)$, is plotted. Total energy is conserved with the error, $\epsilon_E(\tau) < 2.5\times10^{-5}$ until $\tau = 30$.}
\label{nonlin_eren}
\end{figure}
Figure~\ref{erar} shows an error found in the particle (or area) conservation, $\epsilon_N :=|N(\tau) -N(0)|/N(0)$, for the case with 8000 nodes/contour, where $N(\tau)$ is a total integral of the particle density defined as
\begin{align}
N(\tau) := \iint \tilde{f'}(x,v,\tau)dxdv,\label{nt}
\intertext{where}
\tilde{f'}(x,v,\tau) := \displaystyle{\sum^{N_m}_{m=1}}\Delta f'_m I[(x,v)\in S_m(\tau)]
\end{align}
with $S_m(\tau)$ denoting the closed polygonal region determined by nodes on the $m$th contour. Errors in the particle conservation, $\epsilon_N$, is less than $10^{-4}$ while increasing in time. 
Generally speaking, the CD method  tends to fail in following the strong deformation of contours with large curvature, because the contours consist of finite straight segments connected by nodes.
\begin{figure}[h]
\centering
\includegraphics[angle=-90,scale=.5]{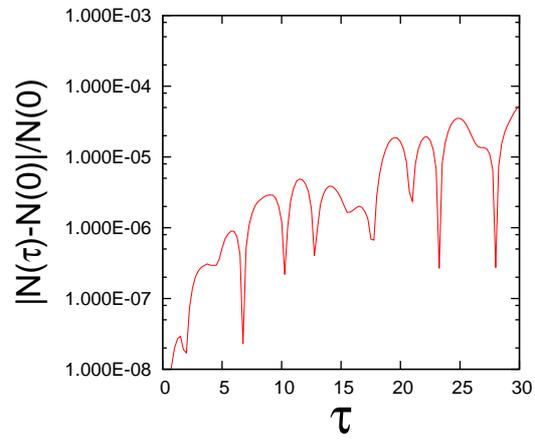}
\caption{Time evolutions of an error found in the particle conservation, $|N(\tau) - N(0)|/N(0)$. $8000$ nodes/contour are employed, where $N(\tau)$ means the integral of the particle density defined in Eq.~(\ref{nt})}
\label{erar}
\end{figure}
\clearpage

\section{Summary and Conclusion}\label{SumAndCon}
We have newly implemented contour dynamics method for the Vlasov-Poisson system with the periodic boundary. The major difficulty in application of the periodic boundary is how to deal with contours when they cross the boundaries. It has been overcome by introducing periodic Green's function defined on the infinite phase space, instead of the Green's function derived for the bounded system with the periodic boundary condition. The new scheme enables implementation without cutting nor reallocating the contours and node points, and accelerates the computational speed.\par
Validity of the CD method for the Vlasov-Poisson system with the periodic boundary is confirmed by comparing the simulation results with the analytical solution for the piece-wise constant distribution function in the linear regime, and by the bench mark test for the linear Landau damping. Nonlinear Landau damping simulation using the CD method successfully reproduces the electron trapping by the Langmuir wave. Soundness of our method is also demonstrated by the energy and particle conservation with errors less than $2.5\times10^{-5}$ and $10^{-4}$, respectively. Improvement of the CD method to reduce the conservation errors remains for future works.\par
Because this paper  focused on the verification of our basic CD scheme for the periodic system, detailed analyses of the physics problem by means of the CD method are remained for future studies. Application of the CD method to a variety of issues in kinetic plasma physics is currently in progress, and will be reported elsewhere.

\clearpage
\section*{Appendix}
Here, we calculate the analytical solution of $a(x,\tau) = {\partial \phi}/{\partial x}$ for the initial distribution function in Eqs~(29) and (30). For a node with the index $m$, $v_m$ satisfies the equation
\begin{align}
&a(x,\tau) =\frac{dv_m}{d\tau}= \frac{\partial v_m}{\partial \tau} + v_m\frac{\partial v_m}{\partial x}.\label{nodeeq1}
\intertext{Eqs.~(\ref{bvppoisson}) and (\ref{linin}) lead to}
& -\frac{\partial a}{\partial x} = 1 - \displaystyle{\sum^{N_m}_{-N_m}}\left(-b_m\right)v_m.\label{pop1}
\end{align}
For the zeroth order, the electron density is assumed to be the same as that of the uniform background ions,$\displaystyle{\sum^{N_m}_{-N_m}}\left(-b_m\right)v^0_m = 1$ (namely, we choose $\Delta v$ to satisfy this relation). Therefore, Eq.(\ref{nodeeq1}) is linearized  as
\begin{align}
& a = \frac{\partial v^1_m}{\partial \tau} + v^0_m\frac{\partial v^1_m}{\partial x}\label{nodeeq2},\\
\intertext{and Eq.~(\ref{pop1}) reads}
&\frac{\partial a}{\partial x} = \displaystyle{\sum^{N_m}_{-N_m}}\left(-b_m\right)v^1_m\label{pop2}.
\end{align}
Assuming $a, v^1_m \propto e^{ikx},$ The Laplace transform of Eqs~(\ref{nodeeq2}) and (\ref{pop2}) give
\begin{align}
&L(a)=-v^1_m(0) + sL\left(v^1_m\right) + v^0_mikL\left(v^1_m\right) \label{lnodeeq2},\\
&-ikL(a) = \displaystyle{\sum^{N_m}_{-N_m}}b_m L\left(v^1_m\right)\label{lpop2},\\
\intertext{where $L\left(f\left(\tau\right)\right): = \int^{\infty}_{0}f(\tau)e^{-s\tau}d\tau$. Thus,}
&L(a)= \frac{1}{D(is)}\displaystyle{\sum^{N_m}_{-N_m}}\left\{\left(-b_m\right)v^1_m(0)\Pi_{l\neq m}\left(is-kv^0_l\right)\right\},\label{La}\\
\intertext{with}
&D(is) = {k\Pi^{N_m}_{m = -N_m}\left(is - kv^0_m\right) +\displaystyle{\sum^{N_m}_{m=-N_m}}\left\{b_m\Pi_{l\neq m}\left(is - kv^0_l\right)\right\}}.
\end{align}
It is known that for a choice of $b_n < 0\;(\forall n\geq1),$ the solutions of $D(is)  = 0$ are  purely real (see Refs.~\cite{Ldawb} and \cite{poptw}). We define $\omega := is$ and $\omega_m:D(\omega_m) = 0 $ with $ kv^0_m < \omega_m < kv^0_{m+1}$ so that $D(\omega_m)$ is written as $D=k\Pi^{N_m}_{m=-N_m}(\omega - \omega_m).$ The inverse Laplace transform of Eq.~(\ref{La}) is
\begin{align}
a &=\displaystyle{\sum^{N_m}_{j=-N_m}} Res\left(L\left(a\right)\left(s\right)e^{st},-i\omega_j\right)\\
& = \displaystyle{\sum^{N_m}_{j=-N_m}} \lim_{s\rightarrow -i\omega_j}\left(s+i\omega_j\right)\frac{1}{k}\frac{\displaystyle{\sum^{N_m}_{-N_m}}\left\{\left(-b_m\right)v^1_m(0)\Pi_{l\neq m}\left(is-kv^0_l\right)\right\}}{\Pi^{N_m}_{m=-N_m}\left(is-\omega_m\right)}e^{st}\\
\intertext{Since $\omega_m = -\omega_{-m},b_m = -b_{-m}$ and $v^0_m = -v^{0}_{-m}$, one finds}
a(x,\tau) &=  Re\left[\displaystyle{\sum_{j>0}}\frac{\displaystyle{\sum_{m>0}}\left\{\left(-b_m\right)\alpha e^{i(kx - \frac{\pi}{2})}\Pi_{n(\neq m)>0}\left(\left(\omega_j\right)^2 - \left(kv^0_n\right)^2\right)\right\}}{k\Pi_{n(\neq j)>0}\left(\omega^2_j-\omega^2_n\right)}2\cos\omega_j \tau\right],
\intertext{where $\omega_j$ satisfies the dispersion relation of}
&1 +\displaystyle{\sum_{m>0}}\frac{2b_mv^0_m}{\omega^2 - \left(kv^0_m\right)^2}=0.%\label{anaD}
\end{align}
\bibliographystyle{unsrt}
\bibliography{ref}
\end{document}